# Imaging emergent exotic quasiparticle state in a frustrated transition metal oxide

Yuita Fujisawa[1], Anjana Krishnadas[1], Chia-Hsiu Hsu[2], Takahito Takeda[3], Sheng Liu[1], Markel Pardo-Almanza[1], Yukiko Obata[1], Dyon van Dinter[1], Kohei Yamagami[4], Guoqing Chang[2], Masaki Kobayashi[3,5], Chang-Yang Kuo[6], Yoshinori Okada[1]

[1]*Quantum Materials Science Unit, Okinawa Institute of Science and Technology (OIST), Okinawa 904-0495, Japan.*
[2]*Division of Physics and Applied Physics, School of Physical and Mathematical Sciences, Nanyang Technological University, 637371 Singapore.*
[3]*Department of Electrical Engineering and Information Systems, The University of Tokyo, Tokyo 113-8656, Japan.*
[4]*Japan Synchrotron Radiation Research Institute, Hyogo, 679–5198, Japan.*
[5]*Center for Spintronics Research Network, The University of Tokyo, Tokyo 113-8656, Japan.*
[6]*National Yang Ming Chiao Tung University, Hsinchu 300, Taiwan.*

The existence of rich Fermiology in anomalous metal phase in exotic superconductors has attracted considerable interests, as exemplified in copper, iron-based, and intermetallic frustrated kagome-based compounds. A common feature in these cases is pseudo-gap opening or long-range lattice/electronic ordering above superconducting critical temperature $T_c$. As yet developed area is the potential existence of exotic Fermiology in superconducting transition metal oxides on a geometrically frustrated lattice. Here, we focus on the spinel oxide superconductor $LiTi_2O_4$, which can be viewed as the hole-doped side of the orbital ordered $3d^1$ Mott system on the Ti-derived pyrochlore frustrated network. By the *in-situ* combination of angle-resolved photoemission spectroscopy (ARPES) and epitaxial thin film growth, we discovered the abrupt flattening of near Fermi energy dispersion below the characteristic temperature $T^*$ ~ 150 K. While the emergent negative thermal expansion below $T^*$ strongly supports a distinct phase at low-temperature, absence of energy gap opening, splitting/folding of bands, nor long-range lattice distortion are seen across $T^*$. We propose that the competition between growing instability towards orbital ordering and its inherent geometric frustration in the Ti-pyrochlore network results in a new quantum state of matter with robust high entropic nature below $T^*$. Our findings collectively point to a unique Fermiology in frustrated three-dimensional transition metal oxides, and its connection to superconductivity below $T_c$ is open as an interesting future challenge. Also, a potential guideline is unexpectedly provided for designing zero thermal expansion metal to develop future solid-state devices.

## Introduction

### Interest in 3d[1] Mott states on a frustrated lattice

A deeper understanding of metallic phases that emerged from Mott insulating states has been an essential challenge in condensed matter physics, which is also connected to developing a universal understanding of exotic superconductors. For example, in cuprates, the pseudogap metal has characteristic momentum dependence, and gap formation becomes the most predominant around saddle point momentum [1,2]. Contrary to the cuprates with predominant $3d_{x2-y2}$ orbital, the orbital-dependent Mottness within $t_{2g}$ state ($3d_{xy}$, $3d_{xz}$, $3d_{yz}$) has been discussed in Fe-based superconductors [3,4]. Above $T_c$, the emergence of nematic ordering with a propagation vector $q = 0$ has been discussed along with band splitting [5]. In these studies, a detailed understanding of the momentum-dependent electronic states near the Fermi energy ($E_F$) has been providing a piece of key information. So far, while most of the research is for square-lattice systems, searching exotic Fermiology in proximity to Mott states on a frustrated lattice has not been well explored experimentally [6]. Although recent intermetallic quasi-two dimensional kagome superconductors with multiple orbital degrees of freedom have gathered considerable interest [7,8], a frustrated lattice with enhanced ionic nature in transition metal oxides is expected to open a new area of research.

### Interest in NTE and frustrated lattice

The transition metal spinel oxides $AB_2O_4$ (**Fig. 1a**) are an exciting platform, which hosts a corner-sharing pyrochlore lattice by element B (**Fig. 1b**). Due to the inherent tetrahedral structure in the pyrochlore network, the arrangement of spin/charge/orbitals can be degenerated and frustrated. Depending on the number of electrons for each site, the frustration of charge (**Fig. 1c**), spin (**Fig. 1d**), and orbital (**Fig. 1e**) has enriched physics in spinel transition metal oxides. While large frustration is an intriguing driving force leading to a rich array of electronic/structural orderings, analogous to spin-liquid states [9,10], an intrinsic absence of the ordering is another exciting concept. For the latter case, regardless of the detailed state, the negative thermal expansion (NTE) is one of the sensitive measures to unveil an emergence of distinct phases of matter [11,12,13,14]. Particularly notable cases are emergent anomalous metallic states adjacent to the $3d^1$ orbital ordered Mott system $MgTi_2O_4$ (**Fig. 1f**) [15,16]. A variety of exotic states have been discovered, including heavy fermions [17], orbital selective spin liquids [18], charge ordering [19], and superconductivity and possible magnetic Weyl semi-metallic states [20,21]. As yet achieved experimental challenge is unveiling exotic momentum-dependent Fermiology in spinel oxides. Although direct imaging of the band structure by angle-resolved photoemission spectroscopy (ARPES) has been long desired [22], the experimental demonstration has yet to be accomplished. This is because of the difficulty in preparing large-scale, high-quality single crystals and obtaining atomically flat surfaces by cleaving, which tends to be uncontrollable [23]. The progress in high-quality epitaxial thin film growth of spinel oxides has opened a promising avenue to circumvent these difficulties.

**Summary**

Here, by establishing an *in-situ* combination of epitaxial thin film growth and ARPES measurements, we show the direct visualization of the band structure in the spinel oxide superconductor LiTi$_2$O$_4$(111)(LTO) film. The electronic state features an effective 3$d^{0.5}$ electronic configuration on a pyrochlore network, which is adjacent to the orbital ordered 3$d^1$ Mott state (**Fig. 1f**). We discovered the existence of a characteristic temperature $T^* \sim$ 150 K, below which a surprising correspondence between the abrupt dispersion kink formation and the NTE is unveiled (**Fig. 1g**). The exotic metal below $T^*$ is a new quantum state of matter, where the development of an orbital ordering is intrinsically suppressed by the geometrical frustration.

## Results and discussions

### ARPES data on LiTi$_2$O$_4$(111)

In LTO, the near $E_F$ bands are dominated by Ti-derived 3$d$ orbitals. The Brillouin zone (BZ) and the band structure from a single orbital tight binding model for the pyrochlore lattice are shown in **Fig. 2a** and **b**, respectively. Similar to two dimensional kagome systems [24,25,26], the expected underlying bare electronic band structure of the pyrochlore lattice hosts flat bands, Dirac points, and saddle points at energy $E_{SP}$ (see **Fig. 2c**) [27]. Four Ti sites exist in this simplest model (two for flat and one for electron and hole branch for the Dirac electron) and $E_F$ for 3$d^{0.5}$ per Ti atom is expected to be close to the $E_{SP}$ (**Fig. 2b** and **c**). **Fig. 2d** shows the Fermi surface (FS) mapping of LTO (111) taken with an incident photon energy of $h\nu$ = 21.2 eV (He-I$\alpha$) at 12 K (see **Methods** for the details). The large FS suggests that the saddle point is close to the $E_F$. The 6-fold symmetric feature is consistent with the fact of the existence of two domains with a 180-degree in-plane rotation and $k_z$ value close to zero (see **Supplementary Note 1** and **2** for the STM topograph and its negligible effect on the main focus of this study). The ARPES intensity maps at 300 K along two different orthogonal directions, $\Gamma_1 - M_1$ and $K_1 - M_1$, clearly demonstrate the hole-like (**Fig. 2e, f**) and electron-like (**Fig. 2g, h**) band dispersion, respectively, confirming the generic feature of a saddle point.

### Comparison with DFT

In **Fig. 3a**, the ARPES dispersion along $\Gamma_1 - M_1$ at 300 K is compared with the density functional theory (DFT) calculation (red curves) that is performed based on the previously reported LTO crystal structure [28]. The DFT calculations confirm that the near-$E_F$ bands are derived from Ti $t_{2g}$ electrons, independent of oxygen and Li-derived states near $E_F$ (see **Supplementary Figure 7a**). As can be seen in **Fig. 3a**, the band hosting the saddle point, coined the $\alpha$-band, can be well reproduced by DFT calculations ($k_z = 0.17$ 1/Å). The solid and dashed lines represent the calculated, nearly degenerate multiple eigenstates derived from the $t_{2g}$ orbitals (see **Supplementary Figure 7b-g**). Since the splitting is much smaller than the experimental momentum resolution, we phenomenologically treat the observed band as a single $\alpha$-band.

**Kink structure near $E_F$**

We confirmed the significant many-body effects on the α-band, which do not appear in non-interacting DFT calculations. A relatively flat band observed at $E - E_F \sim$ -100 meV at 12 and 300 K (denoted as the β-band in **Fig. 3a** and **b**,) is suggested to be the surface state by our DFT calculations (see **Supplementary Note 4**). While a sudden broadening of the momentum distribution curve (MDC) below $E - E_F$ = -200 meV is seen (see black arrows in **Fig. 3a** and **b**), this high energy many-body effect is less dependent on temperature. As the most important many-body effect, the kink structure emerges at 12 K with an energy ($E_{kink}$) of approximately -50 meV (**Fig. 3b**). Notably, this near $E_F$ kink structure is absent at 300 K (**Fig. 3a**). Regardless of temperature, we stress that the absence of the gap opening is confirmed within our experimental resolution (see **Fig. 3c, d**).

**Coupling constant λ for quantitative analysis**

For the quantitative analysis of the kink structure at $E_{kink}$, we phenomenologically introduce the coupling constant λ as the degree of an electron-boson coupling. As a well-accepted methodology [29,30], we define $\lambda = \frac{v_0}{v_r} - 1$ where $v_0$ and $v_r$ are the group velocities of the bare and the dressed electrons. **Figure 3f** represents the energy-momentum dispersion relation, which is obtained from the peak positions of the MDC from **Fig. 3e**. Due to the matrix element effect, the estimation of λ from data with a smaller $k_F \sim$ 0.6 1/Å is more reliable. The extracted λ ~ 1.4 is comparable to other transition metal oxides [31]. Importantly, the kink at $E - E_F \sim$ -50 meV is absent at room temperature, suggesting the presence of a characteristic temperature at which the kink starts to appear.

**The emergence of the characteristic temperature $T^*$**

To unveil the characteristic temperature quantitatively, we performed temperature-dependent ARPES experiments. **Fig. 4a** and **b** show ARPES intensity maps and traces of the dispersion at various temperatures. Since λ is found to be independent of $k_F$ within our experimental resolution (see **supplementary Note 5** for the detail), we choose the same cut as **Fig. 3a** and **b** to evaluate λ. As the temperature decreases (from right to left in **Fig. 4a** and **b**), the kink structure becomes prominent. The temperature variation of the $k_F$ extracted from the MDC curves is confirmed to be negligible, which excludes the extrinsic surface carriers modification (**Fig. 4c** and **d**). From the temperature evolution of λ, the existence of a characteristic temperature of $T^* \sim$ 150 K is found, below which an abrupt increase in electron-boson coupling is observed (**Fig. 4e**).

**$T^*$ as a bulk characteristic**

Before discussing the microscopic picture, we first confirm the bulk nature of $T^*$. For this purpose, we carefully investigated the temperature dependence of an X-ray diffraction (XRD) pattern. Note that our film is fully relaxed from epitaxial strain from the Nb-doped $SrTiO_3$ (STO) substrate, and the overall lattice keeps a cubic symmetry regardless of the temperature (see also **Supplementary Figure 2** and

10). **Figure 5a** shows the XRD pattern at room temperature and well-separated peaks of LTO (444) and STO (222) are observed. From the temperature dependence of the isolated LTO (444) peak (**Fig. 5b**), it is evident that no peak splitting is seen within our experimental resolution. This suggests the absence of a clear structural transition, such as the cubic-to-tetragonal transition previously reported for other spinel oxides [15,16,32]. Nevertheless, we found an intriguing temperature dependence. **Figure 5c** shows the temperature dependence of the normalized lattice spacing defined as $\Delta d_{hkl} = \{d_{hkl} - d_{hkl}(300\text{ K})\}/d_{hkl}(300\text{ K})$ for both LTO (444) and STO (222) (see **Supplementary Figure 11** for the raw data). While two different diffractions show the same trend above 150 K, $\Delta d_{444}$ of LTO becomes minimum around 150 K. Therefore, striking evidence of emergent negative thermal expansion (NTE) below $T^*$ ~150 K has been unveiled. This strongly supports the characteristic distinction across $T^*$ as an intrinsic bulk nature (see **Fig. 1g**). To show the excellent agreement between the estimated $T^*$ by surface-sensitive ARPES and bulk-sensitive XRD, the temperature evolution of λ is overlaid in **Fig. 5c**.

**New quasi-particle state**

We stress that the observation of a robust gapless Fermi surface across $T^*$ in LTO is a non-trivial fact. The spinel structure is known to host frustrated, and therefore, flexible lattice, which promotes a variety of instability towards electronic/lattice long-range orderings and phase transitions [15,16,32]. By increasing the number of electrons from $3d^{0.5}$ to $3d^1$, a recent experiment shows that the electronic instability toward orbital ordering is enhanced systematically [33]. Furthermore, it is worthwhile to point out that the saddle point is seen very close to $E_F$ (**Fig. 2e-h**). While the existence of a saddle point at $E_F$ often leads to a phase transition with gap opening at $E_F$, this is not the case for LTO. While the microscopic physics of the transition across $T^*$ in LTO is not fully clarified at this moment, a major role of inherent frustration can be naturally speculated. We propose that inherent frustration in the Ti-pyrochlore network plays an important role to realize the robust existence of a gapless exotic metal, where the appearance of long-range electronic/lattice ordering is intrinsically suppressed. While the kink energy is similar to the phonon energy in LTO [34], we believe that the existence of a characteristic temperature $T^*$ associated with the emergent NTE strongly suggests a new type of quasiparticle state below $T^*$ beyond the ubiquitous electron-phonon coupling phenomenology. It is interesting to point out that a phenomenological analogy can be seen in a Weyl semimetal, where a gap opening is absent regardless of the existence of a van-Hove singularity at $E_F$, leading to enhanced electron-boson coupling without long-range ordering [35].

**Connection to SC**

Finally, contrasting characteristics to anomalous metal states in other quasi-two dimensional exotic superconductors are discussed. The absence of a gap opening across $T^*$ in LTO is a clear distinction from the partial gap opening for the pseudogap state in cuprates. The absence of band folding and gap opening across $T^*$ in LTO can be partially understood by $q$ = 0 ordering as observed in Fe-based superconductors [5]. However, the absence of clear band splitting (along the energy direction) by lifting orbital degeneracy in LTO is distinct from the case of Fe-based superconductors. While the recent boost of interest in kagome superconductors provides an interesting analogy of the nematic transition with ill-defined lattice distortion, long-range charge density wave transition has been clearly observed [7,8]. Therefore, the absence of a gap opening or long-range phase transition above $T_c$ is unique nature of LTO, compared to copper, iron-based, and intermetallic frustrated kagome based compounds. Overall, our findings collectively point to the existence of a unique Fermiology in quasi-three-dimensional frustrated transition metal oxides, which has not been recognized so far. Unveiling a connection between the emergent anomalous metal below $T^*$ and superconductivity with relatively high $T_c$ is an intriguing future challenge. Moreover, the combination of *in-situ* single particle spectroscopies with epitaxial thin film growth, as employed in this study, has opened a playground for visualizing the rich physics in tunable hetero-interface in correlated spinel/pyrochlore systems [36,37].

## Summary/Impact

In summary, although the emergent NTE below $T^*$ strongly support a distinct phase at low-temperature, neither energy gap opening, band splitting/folding, nor long-range lattice distortion are seen in LTO. We propose that the competition between growing instability towards orbital ordering and its inherent geometric frustration in the Ti-pyrochlore network leads to a new quantum state of matter with robust highly entropic nature below $T^*$. The impact of our finding is not limited to fundamental interests of unique Fermiology in frustrated metals, but also to potential practical aspects. Regarding the recent dramatic downsizing of device length scale and growing requirement of atomic precise hetero interfacial design, for example, our finding can be useful as a guideline to make robust metallic electrodes with zero (or tunable) thermal expansion.

## Methods

**Film growth and *ex-situ* characterization**

Epitaxial thin films were grown using pulsed laser deposition (PLD) [38,39,40,41,42]. We used 0.05 % Nb-doped SrTiO$_3$ (STO) as substrate to demonstrate the negligible interfacial effect (see **Supplementary Figure 2** detailing the negligible epitaxial strain effect). The details of the film growth conditions are similar to the previous study [41]. After the *in-situ* STM/STS and ARPES measurements, the films were characterized by *ex-situ* X-ray diffraction (XRD) for structural analysis. $T_c$ is estimated from magnetization measurements (see **Supplementary Figure 4**). All film characterizations indicate that the quality of our films, including atomic scale flatness, is comparable to the highest standard reported in the literature [41,42]. The surface morphology of the film was also characterized by *in-situ* scanning tunneling microscope (STM). An electrochemically polished tungsten wire is used as an STM tip followed by *in-situ* bombardment, gentle poking, and voltage pulses on a surface of a single crystal gold. STM topographs are taken in constant current mode. See **Supplementary Note 1** for more details.

**In-situ spectroscopy**

ARPES measurements were conducted on samples transferred in pristine condition from the adjacent, UHV-connected PLD system, i.e., without ambient exposure or capping layer [43,44]. The DA30HL (Scienta Omicron™) electron analyzer was used to study photo-excited electrons with an instrumental energy resolution < 10 meV. Both He-I ($h\nu = 21.2\ eV$) and He-II ($h\nu = 40.8\ eV$) are used to examine the effect of probing different $k_z$ values. See **Supplementary Note 3** for the data taken with He-II.

**DFT calculations**

DFT-based first-principles calculations were implemented in the Vienna ab-initio simulation package (VASP) [45,46] using Projector-augmented-wave (PAW) pseudopotentials [47]. The generalized gradient approximation (GGA) in the Perdew-Burke-Ernzerhof (PBE) form [48] was used to treat exchange-correlation effects. An energy cut-off of 500 eV was used throughout the calculations. The structures were optimized until the residual forces were less than $10^{-3}$ ($10^{-2}$) eV/Å and the self-consistency criterion for convergence was set to $10^{-6}$ ($10^{-5}$) eV for the bulk (surface) system. Γ-centered Monkhorst-Pack [49] grids of size $9 \times 9 \times 9$ were used for the bulk of the rhombohedral lattice ($a = b = c = 5.986\ Å, \alpha = \beta = \gamma = 60°$). Additionally, the grids of $9 \times 9 \times 1$ were used, and a vacuum layer of 15.0 Å was added along the *z*-direction to eliminate the interaction due to periodic boundary conditions in the surface system. SOC effects were included in our band structure calculations. For comparison between the APRES and the DFT calculation, a shift of 0.072 eV is implemented, which is negligibly small within the primary purpose of this study.

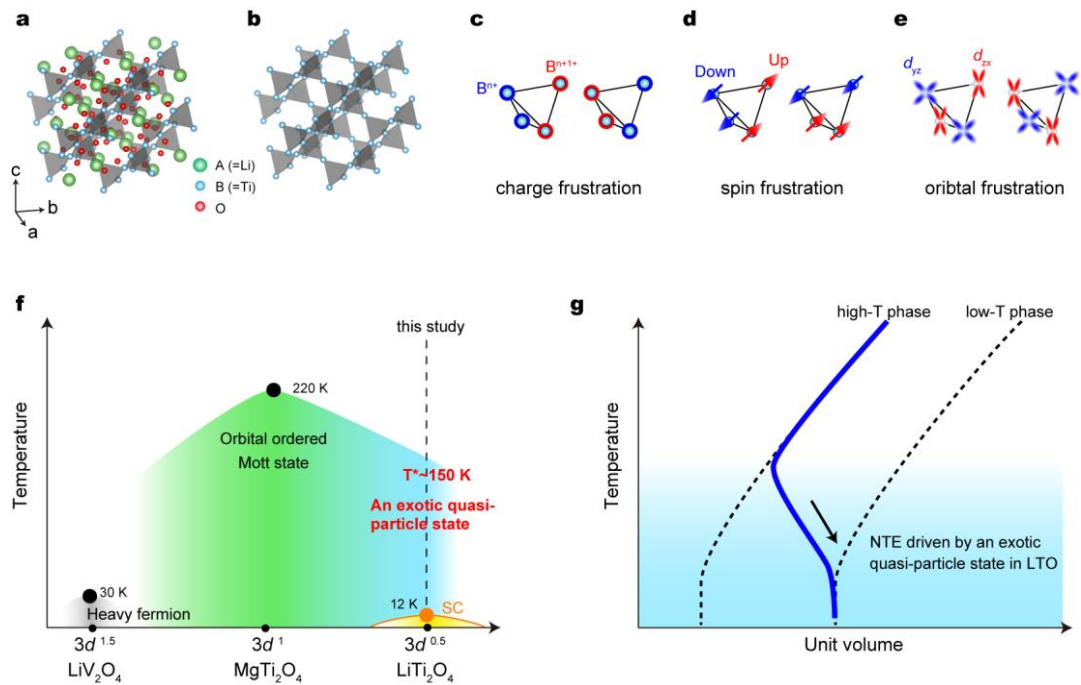

**Fig. 1| Spinel oxides with geometrical frustration and their phase diagram.**

**a,** Crystal structure of LiTi$_2$O$_4$.

**b,** Pyrochlore lattice of Ti atoms.

**c-e,** Schematic representation of the frustrated electronic degrees of freedom, such as charge (c), spin (d), and orbital (e).

**f,** The phase diagram of spinel oxides near $d^1$ electronic configuration. Rich electronic phases are displayed.

**g,** Schematic representation of the mechanism of negative thermal expansion (NTE). The NTE usually occurs when a phase transition/crossover behavior is observed from a high temperature phase with smaller unit volume to a low temperature phase with larger unit cell.

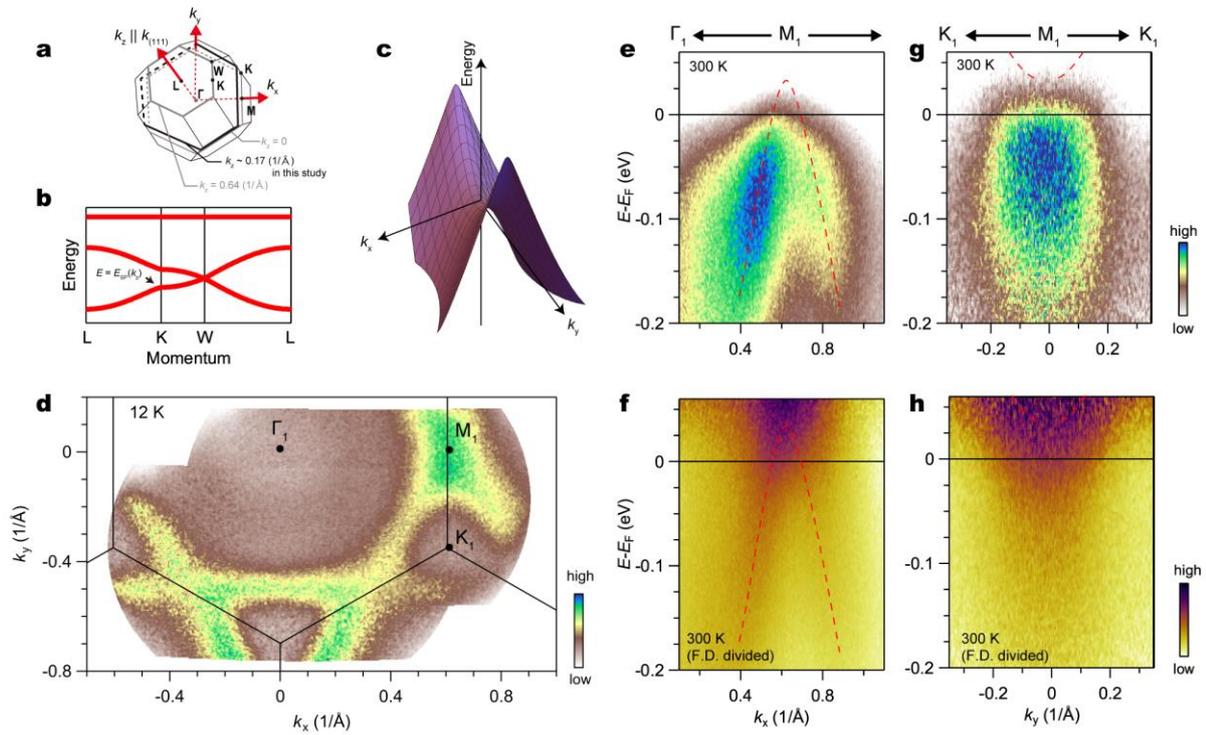

**Fig. 2| Overall band structure of LTO (111).**

**a,** Three-dimensional Brillouin zone. In this study, $k_z$ is defined parallel to the (111) direction.

**b,** Tight-binding calculation for the pyrochlore lattice. The black arrows show the saddle point at energies of $E_{SP}$ which depend on $k_z$.

**c,** Schematic representation of a saddle point.

**d,** Fermi surface taken with the He-I light source. Note that the black lines serve as guides for the eyes. The labels of $\Gamma_1$, $M_1$, and $K_1$ are defined based on the obtained Fermi surface.

**e-h,** Band dispersion near the Fermi energy taken at room temperature to search for the saddle point. ARPES mappings are shown along $\Gamma_1 - M_1$ (e,f) and $K_1 - M_1$ (g,h). Raw data are shown in (e) and (g), while the Fermi-Dirac function-divided images are shown in (f) and (h).

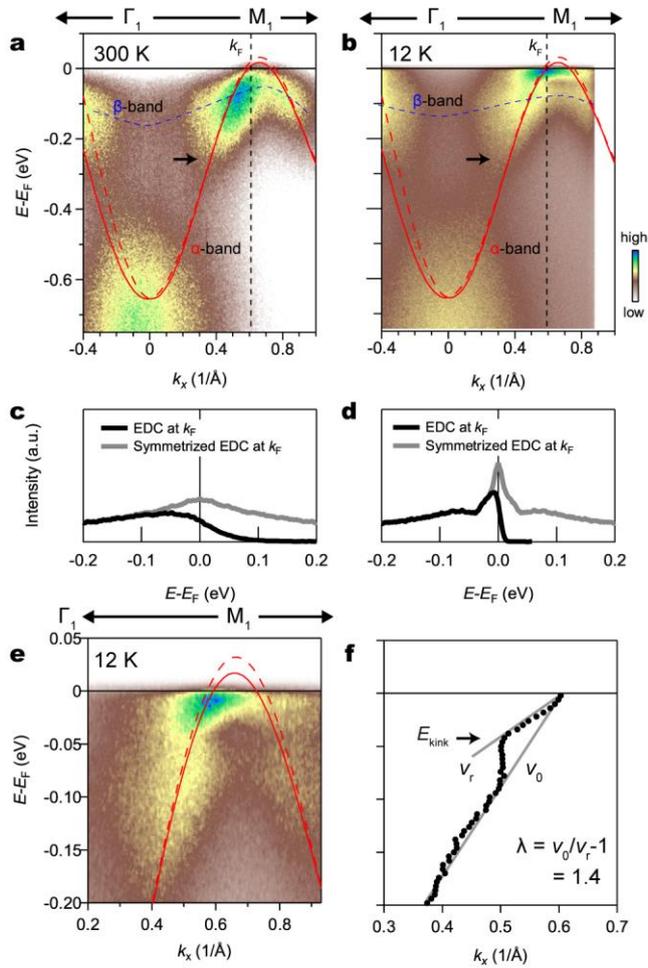

**Fig. 3| Temperature variation of the band structure in LTO (111).**

**a,b,** ARPES dispersion along $\Gamma_1 - M_1$ taken at 300 K and 12 K, respectively. Two red lines (solid and dashed) are DFT-derived dispersions taken at $k_z = 0.17\ (1/\text{Å})$.

**c,d,** Energy distribution curves and their symmetrized curves taken at 12 K and 300 K, respectively.

**e,** A magnified ARPES image near $E_F$ at 12 K. The red curves are the same DFT dispersions as in a and b.

**f,** Trace of the band dispersion. The coupling constant is estimated from the equation in the inset.

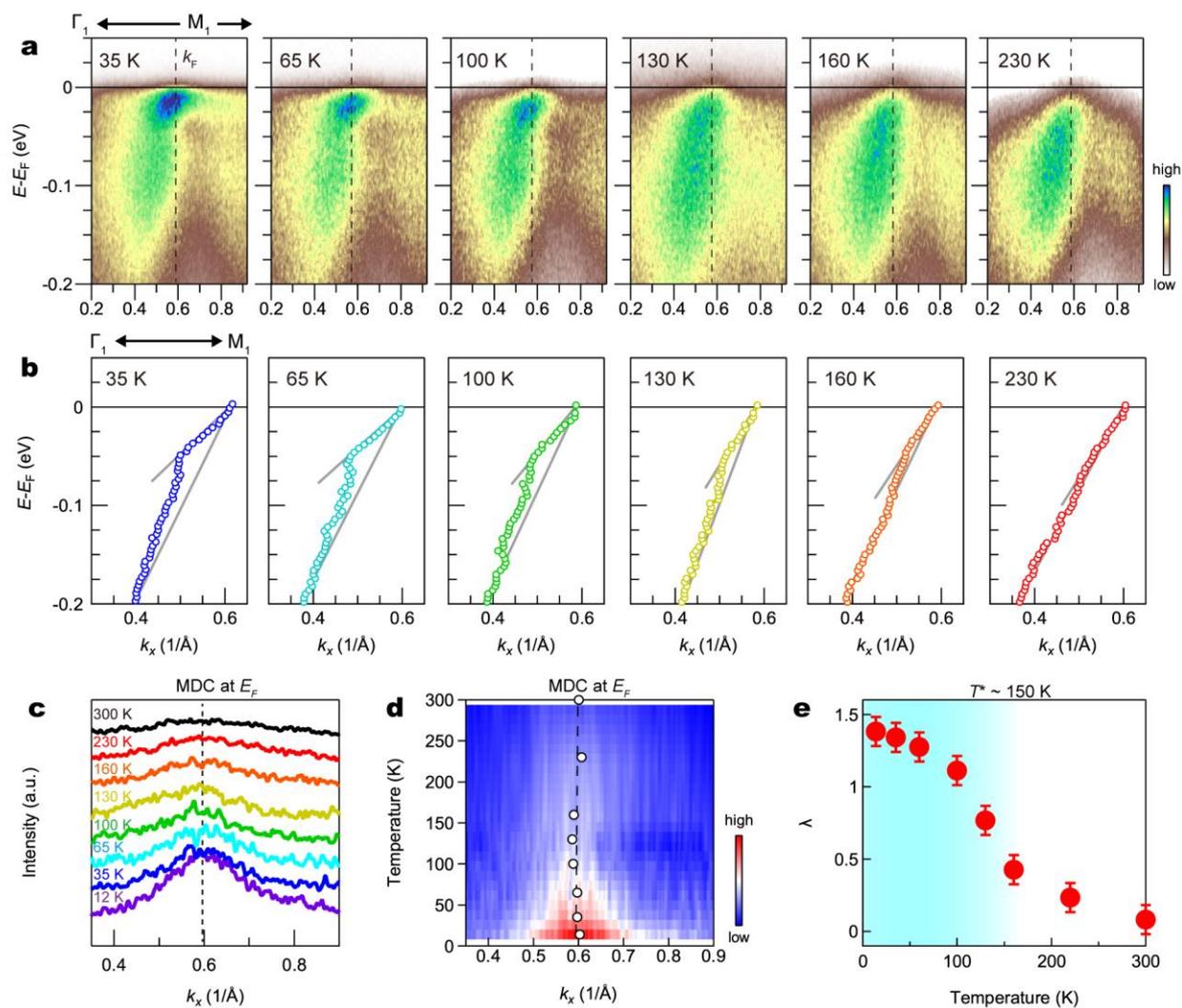

**Fig. 4| Systematic investigation of the band structure as a function of temperature.**

**a,b,** ARPES raw data taken along $\Gamma_1 - M_1$ at various temperatures (a) and trace of the dispersion (b).

**c,** MDCs at $E_F$ taken at various temperatures.

**d,** Color plot of MDCs in (c) as a function of temperature.

**e,** Temperature dependence of the coupling constant λ showing a sudden increase around 150 K.

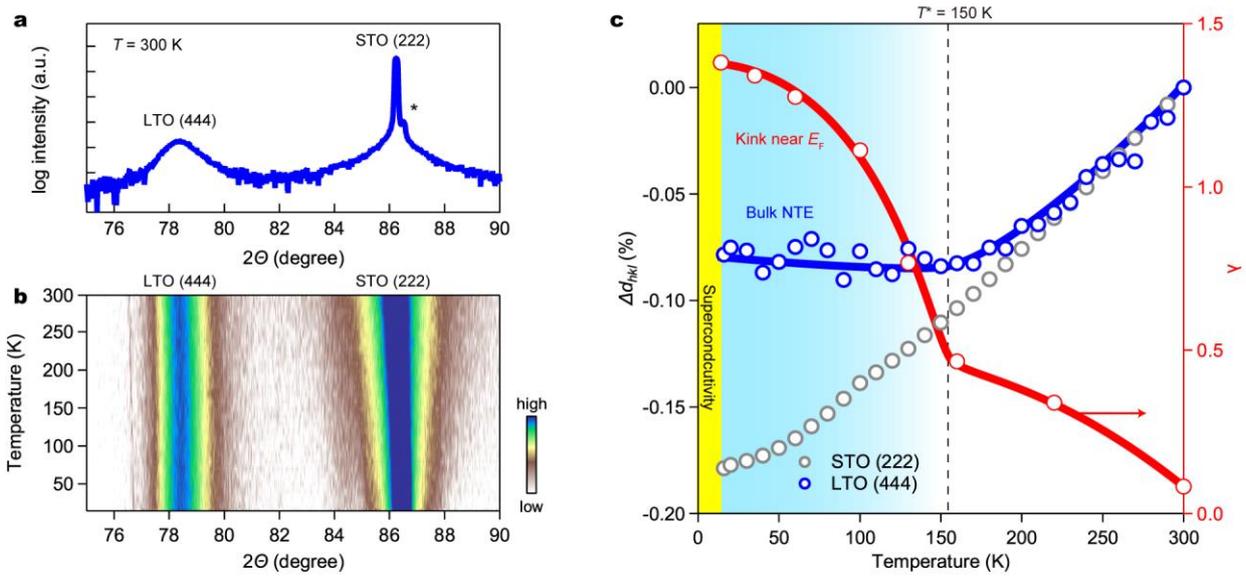

**Fig. 5| Detection of a bulk anomaly at 150 K by XRD.**
**a,** $2\theta/\theta$ scan of an LTO(111) film on an STO substrate taken at room temperature. * denotes a diffraction peak of STO (222) due to residual $K_{\alpha2}$ radiation.
**b,** Color plot for the temperature evolution of the $2\theta/\theta$ scan of (a).
**c,** Temperature dependence of $d_{hkl}$ normalized to the value at room temperature for LTO(444) and STO(222) ($\Delta d_{hkl} = \{d_{hkl} - d_{hkl}(300\,\text{K})\}/d_{hkl}(300\,\text{K})$). The blue curve represents a guide to the eye. The coupling constant λ derived from ARPES is also shown on the right axis to demonstrate correspondence with each other.